\documentstyle[prl,aps]{revtex}

\begin{document}
\draft

\twocolumn[\hsize\textwidth\columnwidth\hsize\csname@twocolumnfalse%
\endcsname

\title{Statistics of S-matrix poles for chaotic systems with broken
time reversal invariance: a conjecture}
\author{
Yan V. Fyodorov$\S,\P$
Mikhail Titov$\P$
and Hans-J\"{u}rgen Sommers$\S$       }
\address{
$\S$Fachbereich Physik, Universit\"at-GH Essen,
D-45117 Essen, Germany
        }
\address{
$\P$  Petersburg Nuclear Physics Institute, Gatchina
188350, Russia }

\date{\today}

\maketitle

\begin{abstract}
 In the framework of a random matrix description
of chaotic quantum scattering the positions of
$S-$matrix poles are given by complex eigenvalues $Z_i$
of an effective non-Hermitian random-matrix
Hamiltonian.
We put forward a conjecture on statistics of $Z_i$
for systems with broken time-reversal invariance and
verify that it allows to reproduce
 statistical characteristics of Wigner time delays
known from independent calculations.
We analyze the ensuing two-point statistical measures
as e.g. spectral form factor and the number variance.
In addition we find the density
of complex eigenvalues of real asymmetric matrices generalizing
the recent result by Efetov\cite{Efnh}.

\end{abstract}

 ]

One of the basic concepts in chaotic quantum scattering is the notion
of resonances, representing long-lived intermediate states to which
bound states of a "closed" system are converted due to coupling to
continua. On a formal level resonances show up as poles of the
$M\times M$ scattering matrix $S_{ab}(E)$ occuring at complex energies
${\cal E}_k\!=\!E_k\!-\!(i/2)\Gamma_k$,
where $E_k$ is called the position
and $\Gamma_k$ the widths of the corresponding resonance, and $M$
is the number of channels open in a given interval of energies.
Recently, advances in computational techniques made available
 resonance patterns of high accuracy to be obtained for realistic
models of atomic and molecular chaotic systems\cite{Blumel}.

As is well-known, universal statistical properties of bound states
in the regime of quantum chaos have their pattern in statistics of
real eigenvalues of large random matrices\cite{Bohigas,AlSi}. The
methods to adjust random matrix description to the case of resonance
scattering in open quantum systems are very well-known since the
pioneering work by the Heidelberg group\cite{VWZ}. The method proved
to be very fruitful and allowed to calculate different
universal characteristics of chaotic scattering, see papers
\cite{FSR,FA} for a thorough discussion of recent developments.

In the framework of this approach the S-matrix poles (resonances)
are just the complex eigenvalues of an effective random matrix Hamiltonian
${\cal H}_{eff}\!=\!H\!-\! i\Gamma$. Here $H$ is a large self-adjoint
$N\times N$ matrix of appropriate symmetry serving to describe the
statistical properties of the {\it closed}
counterpart of the scattering system under consideration.
The $N\!\times\! N$ matrix $\Gamma$ is
to describe a possibility of transitions from the states described
by $H$ to the outer world via $M$ open channels and is
related to the $N\!\times\! M$ matrix $W$ of transition amplitudes
as $\Gamma\!=\!\pi WW^{\dagger}$.
Such a form is actually dictated by
the requirement of $S-$matrix unitarity and ensures that all the
$S-$matrix poles are in the lower half-plane of complex energies
as required by causality.

In spite of quite substantial
 analytical \cite{Sok,Haake,FSR} and numerical \cite{num}
work on properties of ${\cal H}_{eff}$ our actual
knowledge of resonance statistics is rather limited.
Apart from the simplest perturbative Porter-Thomas
 treatment \cite{Porter} ( as well as its more advanced
 variants \cite{pt})
the results available on statistics of $S-$ matrix poles are: (i) the
joint probability density of all resonances for the system with one
open channel and unbroken time-reversal invariance (TRI)\cite{Sok}, (ii) the
mean density
of S-matrix poles in the complex plane for large number of open
channels $M\!\sim\! N\!\gg\! 1$\cite{Haake}, (iii)
the mean density of S-matrix poles for arbitrary $M\!\ll\! N$
for the case of broken TRI \cite{FSR}.
In particular, no information about two-point correlations between
different poles is available to our best knowledge.

The situation improves drastically if one replaces the physically
motivated matrix $\Gamma$
introduced above by an (unphysical) $N\!\times\! N$ Hermitian matrix
$A$ with independent, Gaussian distributed elements and considers
$H$ to be taken from the Gaussian Unitary Ensemble (GUE), thus
 restricting attention to the systems with broken TRI.
When the variance
of both $H$ and $A$ coincide, such an ensemble was studied long ago by
Ginibre\cite{Gin}, who managed to find all the correlation functions of
the eigenvalues in the complex plane.  They turned out to be quite
different from those known for the self-adjoint Gaussian random
matrices with real eigenvalues, studied by Wigner,Dyson,Mehta and
others\cite{Bohigas}.

At the same time it is clear that reducing the variance of $A$ as
compared to that of $H$ drives the ensemble towards GUE.
 The existence of a nontrivial regime
of {\it weak non-Hermiticity} was recognized in our preceding works
\cite{FKS1,FKS2}.
This regime occurs when $\mbox{Tr} A^2\!\sim\! \frac{1}{N}TrH^2$
in the limit of large $N$. Under this condition the
imaginary part $Y_k$ of a typical complex eigenvalue
$Z_k\!=\!x_k\!-\!iY_k$
is comparable
with the mean {\it separation} $\Delta\!=\!\langle
\!x_k\!-\!x_{k+1}\!\rangle\! \sim\! \frac{1}{N}$ between
neighboring eigenvalues along the real axis.
 Exploiting the method of orthogonal polynomials
we demonstrated \cite{FKS2} that all statistical properties of
${\cal H}\!=\!H\!-\!iA$ in this regime can be described in
terms of a
kernel $K(Z_1,Z_2)$ depending on two complex coordinates
$Z_{1,2}\!=\!x\!\pm\!{\omega}/{2N}\!-\!i{y_{1,2}}/{N}$.
In particular, the mean eigenvalue density $\rho(Z)=\langle
\>\sum_k\delta^{(2)}\!(Z\!-\!Z_k)\rangle$ is given by
$\rho(Z)=K(Z,Z)$
and the two-point {\it cluster function} ${\cal Y}_2(Z_1,Z_2)$
defined via the relation:
\begin{equation}\label{clust}
\langle \rho(Z_1)\rho(Z_2)\rangle_c\!=\!\langle \rho(Z)\rangle
\>\delta^{(2)}\!(Z_1\!-\! Z_2)\!-\!{\cal Y}_2(Z_1,Z_2)
\end{equation}
 is given by
${\cal Y}_2(Z_1,Z_2)=|K(Z_1,Z_2)|^2$.

Unfortunately, such a detailed information is of no direct use
for the physically motivated case of chaotic scattering.
Nevertheless, the insights provided by the Gaussian case
combined with the existent knowledge of the resonance statistics
allowed us to put forward a well-grounded conjecture about
statistics of complex eigenvalues of weakly non-Hermitian
 ensemble of the type $H\!-\! i\Gamma$ for {\it any} given
Hermitian matrix $\Gamma$. This issue is the content of the main part
of the present paper.

Before formulating the conjecture, let us recall that the GUE
ensemble is an {\it invariant} one, i.e. its statistics is the same
independently on the basis chosen. Therefore, one always can go
to the eigenbasis of the matrix $\hat{\Gamma}$ and consider it
to be diagonal with eigenvalues $\gamma_i, \,\, i=1,...,N$.
In what follows we find it convenient to characterize
the matrix $\hat{\Gamma}$ by the following function:
\begin{equation}\label{f}
f_{\Gamma}(z)=\sum_{i}\ln{\left(1\!+\!\frac{z}{\pi\nu(x)g_i}\right)}
\end{equation}
where $g_i=\frac{1}{2\pi\nu(x)}
(\gamma_i+\gamma_i^{-1})$ and
$\nu(x)=\frac{1}{2\pi}\sqrt{4-x^2}$ stands for the Wigner
semicircular density of
real eigenvalues of the Hermitian part $\hat{H}$ of the matrices
${\cal H}$.

Now we put forward the following {\bf conjecture}:\\
Suppose the function $f_{\Gamma}(z)$ defined above has a finite
limit when $N\!\to\! \infty$. Then
the statistics of eigenvalues $Z_i$
of the corresponding
 almost Hermitian ensemble ${\cal H}_{eff}$ in the limit $N\!\gg\! 1$  is
completely determined by the kernel
\begin{eqnarray}\label{kern}
&K&(Z_1,Z_2)\!=\!\frac{N^2}{4\pi^2}
e^{i\frac{x}{2}(y_2-y_1)}\!\!\!\!\!\!\!\!
\int\limits_{-\pi\nu_{sc}(x)}^{\pi\nu_{sc}(x)}
\!\!\!\!\!\!\!\!du
e^{-u(y_1+y_2)+iu\omega+f_{\Gamma}(u)}
\\
\nonumber
&\times&\!\!
\left(
\int\limits_{-\infty}^{\infty}\!\!\!\!dk_1
e^{-ik_1y_1-f_{\Gamma}(-ik_1/2)}\!\!
\int\limits_{-\infty}^{\infty}\!\!\!\!dk_2
e^{-ik_2y_2-f_{\Gamma}(-ik_2/2)}\right)^{1/2}
\end{eqnarray}

In particular, the mean eigenvalue density $\rho(Z)=\langle
\ \sum_k\delta^{(2)}(Z\!-\!Z_k)\rangle$ is given by
$\rho(Z)\!=\!K(Z,Z)$
and the two-point cluster function is
${\cal Y}_2(Z_1,Z_2)\!=\!|K(Z_1,Z_2)|^2$.

Let us now systematically veryfy the compatibility of our conjecture
with the known properties of
almost-Hermitian matrices of various types.

The simplest test is to make sure that for a Gaussian $\hat{\Gamma}$
such that $Tr{\Gamma}^2\!\sim\! \frac{1}{N} Tr H^2$ we are
back to results proved in \cite{FKS2}. Indeed, for this case
a typical eigenvalue $\gamma_i\sim {N^{-1/2}}\ll 1$, hence
$g_i^{-1}\approx 2\pi\nu(x)\gamma_i\!\ll\! 1$ and in the limit of large
$N$ one can expand $f(z)$ in a series. The first term (proportional to
$z$) vanishes in the limit $N\!\to\! \infty$ because of the symmetry of
the distribution of eigenvalues of ${\hat{\Gamma}}$ around zero. Thus, the
leading term turns out to be proportional to $z^2$. The
corresponding Gaussian integrals over $k_{1,2}$ in Eq.(\ref{kern})
can be performed exactly, the resulting kernel reproducing that found
in\cite{FKS2}.

Let us now consider the mean eigenvalue density
$\rho(Z)\!=\! K(Z,Z)$. The "physical" case
$\Gamma\!=\! \pi WW^{\dagger}$
with a finite number $M$ of open channels (i.e, with a finite number $M$
of {\it positive} non-zero eigenvalues $\gamma_i,\,i=1,..,M$)
was considered earlier in \cite{FSR}. The result coincides with that
following from Eq.(\ref{kern}).

Actually, one can easily adopt the methods used in \cite{FSR} to satisfy  
oneself that the validity of the corresponding
expression is not restricted by the case of positive $\gamma_i$, but rather
extends to an arbitrary set of eigenvalues. This
fact provides a proof of our conjecture on the level of
the mean eigenvalue density.

Let us now show that our conjecture survives a much more stringent
test on the level of two-point correlations.
 For this purpose let us invoke the notion of the so-called
(energy dependent) {\it Wigner time delay} defined in terms of
resonance positions $E_k$ and widths $\Gamma_k$ as (see \cite{FSR} for more
details):
\begin{equation}\label{td}
\tau_w(E)=\frac{1}{M}\sum_{k}\frac{\Gamma_k}{(E-E_k)^2+\Gamma_k^2/4}
\end{equation}
Using this expression it is easy to relate the correlation function
$\langle \tau_W(E_1)\tau_W(E_2)\rangle_c$ of the Wigner time delays at two
different energies
$E_{1,2}\!=\!E\!\pm\!\Omega/2$
to the two-point correlation function
$\langle \rho(Z_1)\rho(Z_2)\rangle_c$
of the densities of $S-$matrix poles in the complex plane
$Z\!=\!x\!-\! iY$.
Considering the energy difference $E_1\!-\!E_2\!=\!\Omega$ to be
comparable with the mean level spacing $\Delta\!=\!1/{\nu(E)N}$ and
exploiting the fact that both the mean density
$\langle \rho(x,Y)\rangle$
and the cluster function ${\cal Y}_2(x_1,x_2,Y_1,Y_2)$
change with $x\!=\!(x_1\!+\!x_2)/2$ on a scale much larger than
$\Delta$, one can perform the $x-$ integration explicitly.
After this it is convenient to pass to the scaled
variables:$$\tilde{\tau_W}=\frac{M\Delta}{2\pi}\tau_w;\>
\tilde{\omega}=\frac{\pi\Omega}{\Delta};\> y=\frac{\pi Y}{\Delta};
\>\omega_x=\frac{\pi\omega}{\Delta},$$ with  $\omega=x_1\!-\!x_2$,
to rescale the cluster function and the density as follows:
$$\tilde{\rho}_E(y)\!=\!\frac{\Delta^2}{\pi}\langle
\rho(E,Y)\rangle;\> \tilde{\cal Y}_2(E,\omega_x,y_1,y_2)\!=\!
\frac{\Delta^4}{\pi^2}
{\cal Y}_2(E,\omega,Y_1,Y_2)$$ and
 to make a Fourier transformation with respect to
$\tilde{\omega}$. These manipulations allowed us to write down
the relation we were looking for in quite a compact form:
\begin{eqnarray}
\label{rel1} \nonumber
&&\frac{1}{\pi}\!\!\!
\int\limits_{-\infty}^{\infty}\!\!\!\!d\tilde{\omega}
e^{i\tilde{\omega}t}
\left\langle\! \tilde{\tau}_W \!
(E\!+\!\frac{\Delta}{\pi}\tilde{\omega})
\tilde{\tau}_W\!(E\!-
\!\frac{\Delta}{\pi}\tilde{\omega})
\!\right\rangle_c \!\! = \!\!
\int\limits_0^{\infty}\!\!\!dye^{-2ty}\tilde{\rho}_E(y)
\\
&&-\int\limits_{-\infty}^{\infty}\!\!\!d\omega_x
\int\limits_0^{\infty}\!\!dy_1\int\limits_0^{\infty}\!\!dy_2
e^{-t(y_1+y_2+i\omega_x)}\tilde{\cal Y}_2(E,\omega_x,y_1,y_2)
\end{eqnarray}

Such a relation between the correlation function of time delays and
two-point cluster function provides a possibility
of the most non-trivial test of our conjecture. Indeed,
both the left-hand
side and the first term in the right-hand side are known
 from independent calculations\cite{FSR}, which allows us to
rewrite the Eq.(\ref{rel1}) for $t>0$ as:
\begin{eqnarray}\label{rel2} \nonumber
&-&\!\int_{-\infty}^{\infty}\!\!d\omega_x\!\!
\int_0^{\infty}\!\!dy_1\!\!\int_0^{\infty}\!\!dy_2
e^{-t(y_1+y_2+i\omega_x)}\tilde{\cal Y}_2(E,\omega_x,y_1,y_2)
\\ &=&
\frac{1}{2}\theta(2-t)
\int_{-min{(t,1)}}^{1-t}\!\!\! d\lambda\prod_{i=1}^M
\frac{g_i+\lambda}{g_i+\lambda+t}
\end{eqnarray}
Although this relation does not provide a possibility to extract the cluster
function in a unique way, it is easy to satisfy oneself that
a direct subtitution of ${\cal Y}_2(Z_1,Z_2)\!=\! |K(Z_1,Z_2)|^2$, with
the kernel taken from Eq.(\ref{kern}) into the left-hand side of
Eq.(\ref{rel2}) produces after rescaling and
integration exactly the right-hand side of Eq.(\ref{rel2}). This fact
provides the strongest support to our conjecture.

Having at our disposal the conjectured form of the cluster function ${\cal
Y}(Z_1,Z_2)$
it is interesting to calculate other related quantities
frequently used in applications, such as
the spectral formfactor and the number variance. For the Gaussian
case such a calculation was performed in \cite{FKS2}. For the physically
interesting case of resonance scattering the formfactor
can be obtained along the same lines and is equal to:
\begin{eqnarray}\label{b}
\nonumber
&b&\!\!(q_1,q_2,k)=
\!\!\!\int\limits_{-\infty}^{\infty}\!\!d\omega_x\!\!
\int\limits_0^{\infty}\!\!dy_1\!\!\int\limits_0^{\infty}\!\!dy_2
e^{2\pi i(q_1y_1+q_2y_2+k\omega)}{\cal Y}_2(Z_1,Z_2)\\
&=&
\frac{N^4}{2}\theta(\nu\!-\!|k|)\!\!\!\!\!\!\!\!
\int\limits_{-(\nu-|k|)}^{(\nu-|k|)}
\!\!\!\!\!\!\!\!dv
\prod_{i=1}^M\frac{(g_i\nu+v)^2-k^2}{(g_i\nu\!+\!v\!-\!iq_1)
(g_i\nu\!+\!v\!-\!iq_2)}
\end{eqnarray}
with $\nu\!\equiv\! \nu(X)$.
The variance of the number of resonances in a strip
$0\!<\!\mbox{Re}Z\!<\! L;\,\,
-\infty\! <\!\mbox{Im}Z\! <\!\infty$
of the width $L\!=\!L_x\Delta$ comparable with $\Delta$ can be expressed in
terms of $b(0,0,k)$, see \cite{FKS2}. In particular,
for the case of equivalent scattering channels:
$g_i\!=\!g,\,\, i\!=\!1,...,M$ it is given by:
\begin{eqnarray}\label{var} \nonumber
\Sigma_2(L_x)&=&L_x -
\frac{1}{\pi^2}
\int_0^{1}{dk}k^{-2}\sin^2{(\pi k L_x)} \\
&\times& \int_{-(1-k)}^{(1-k)} dv
\left[1-{k^2}(g+v)^{-2} \right]^M
\end{eqnarray}
where we have used $\pi\nu(0)\!=\!1$.
Let us discuss
for simplicity its typical features for the symplest case of only one open
channel $M\!=\!1$. We are actually interested in deviations of the
number
variance from its value known for Hermitian GUE matrices. One finds:
\begin{eqnarray}\label{var1}
\delta\Sigma(L_x)&=&\Sigma_2(L_x)-\Sigma_2^{(GUE)}(L_x) \\ \nonumber
&=&\frac{2}{\pi^2}\int_0^{1}dk\sin^2(\pi k L_x)
\frac{(1-k)}{g^2-(1-k)^2}
\end{eqnarray}
Usually, one is interested in the behaviour of
the number variance for $L_x\!\gg\! 1$. Then, for any $g\!>\!1$ one
finds that the difference $\delta\Sigma(L_x)$ tends to a constant
value $\frac{1}{2\pi^2}\ln{[g^2/(g^2\!-\!1)]}$. The so-called
"perfect coupling" case $g\!=\!1$ is known for being specific: e.g.
the resonance density shows at $g\!=\!1$ the powerlaw
behaviour\cite{FSR}.  On the level of number variance such a
specificity is reflected in a logarithmically growing
difference: $\delta\Sigma(L_x)\!\propto\! \ln{L_x}$.

Finally, the knowledge of the cluster function
allows one to find the small-distance behaviour of the so-called
nearest-neighbour distance distribution $p(Z_0,S\!\ll\!\Delta)$, see
e.g.\cite{FKS2}.
As expected\cite{num}, the leading term for $S\to 0$ turns out to be always cubic:
$p(Z_0,S)\!\sim\! S^{3}$, as long
as the system is open: $g\!\sim\! 1$. However, it is interesting to note
that for very weakly open systems: $g\!\gg\! 1$ the cubic law is
modified in such a way, that in a parametrically
large region $g^{-1}\!\ll\! S/\Delta\!< \mbox{Im}Z_0/\Delta $ the behavour becomes
$p(Z_0,S)\!\sim\! S^{5/2}$ (cf.\cite{FKS2}).  Unfortunately, the
density of complex eigenvalues is exponentially small under the condition
$\mbox{Im}Z_0/\Delta\!\gg g^{-1}$,
the fact hindering a
possibility to observe such an unusual regime numerically.

Let us now turn our attention to another class of random matrices with
complex eigenvalues which attracted much attention
recently. Namely, we consider the ensemble of weakly asymmetric
matrices with real elements. Such matrices can always be presented
in the form $H\!+\! A$, with $H$ being real symmetric (hence, taken from
GOE) and $A$ being a real antisymmetric: $A_{ij}\!=\!-A_{ji}$ such
that $N\mbox{Tr}A^2
\!\sim\! \mbox{Tr}H^2$ in the limit of large $N$.  The
case of matrices $A$ with independent, identically distributed Gaussian
entries was studied by various authors and by different
methods\cite{Efnh,Som,Edel}. In particular, the following
unusual property was detected numerically in\cite{Som} and
proved analytically in \cite{Edel,Efnh}: the {\it finite fraction}
of eigenvalues stays on the real axis even for
$A\!\ne\! 0$. This fact
should be contrasted with the corresponding property of earlier
discussed weakly non-Hermitian matrices whose eigenvalues with
probability unity have a finite imaginary part as long as ${\hat{\Gamma}}\ne
0$.  More recently, the interest to the ensemble of slightly asymmetric
real matrices arose after the work by Efetov\cite{Efnh}, who discovered
its relation to an interesting problem of motion of vortices in
disordered superconductors with columnar defects\cite{Nels}.  Efetov
calculated explicitly the mean density of eigenvalues for the Gaussian
$A$. Shortly after Halasz et al\cite{QCD} discovered that  Efetov's
result describes also the density of real eigenvalues of some matrices
appearing in a random matrix approach to the problem of spontaneous
breaking of chiral symmetry in QCD. An interesting feature
 was that the perturbation considered in \cite{QCD} forcing the eigenvalues
to leave the real line
 was  not at all random. Translating these results
to the ensemble of random real non-symmetric matrices it is natural to expect
that for antisymmetric  perturbations of the type
$A\!=\!\left(\begin{array}{cc} 0&\mu {\bf 1}\\-\mu{\bf
1}&0\end{array}\right)$, with a constant $\mu$ being of the order of $1/\sqrt{N}$
 Efetov's formula should still provide the correct eigenvalue density.

This fact motivated us to reconsider the problem
of calculation of the mean eigenvalue density
for a general fixed real antisymmetric matrix $A$ as it is
 done above for the case of almost-Hermitian matrices. Invoking again the
arguments of the rotational invariance, it is enough to consider
the matrices $A$ of the following structure:
$A=\mbox{diag}(A_1,...,A_N)$, with each block $A_i$ being $2\times 2$
matrix of the form $A_i\!=\!\left(\begin{array}{cc} 0&\mu_i\\-\mu_i
&0\end{array}\right)$, since an arbitrary antisymmetric $A$
can be reduced to such a form by an orthogonal rotation.
The density of complex eigenvalues for the matrix $H+A$ can be found
by a straightforward modification of the calculation presented in
\cite{Efnh}. Introducing the scaled variable $y=\pi\nu(X)N\mbox{Im}Z$
and rescaling the density correspondingly:
$\rho_X(y)\!=\!\langle \rho(Z)\rangle/(N\pi\nu(X))^2$, one finds:
\begin{eqnarray}\label{as}
\rho_x(y)&=&\delta(y)\int_0^1\! du e^{\frac{1}{2}[f_A(\pi\nu u)+f_A(-\pi\nu u)]}\\
\nonumber  & +& \frac{1}{2\pi}
\int_0^1\!du u \sinh{(|y|u)}e^{\frac{1}{2}[f_A(\pi\nu u)+f_A(-\pi\nu u)]} \\
\nonumber &\times&
\int_{|y|}^{\infty}\!\!ds
\int_{-\infty}^{\infty}\!\!dk
e^{-iks-\frac{1}{2}[f_A(i\pi\nu k)+f_A(-i\pi\nu k)]}
\end{eqnarray}
where the function $f_A(z)$ is given again by Eq.(\ref{f}) with
$\gamma_i$ replaced by $\mu_i$.

From the derived expression one immediately infers that if
a typical $\mu_i$ is of the order of $N^{-1/2}$, the function
$f_A(u)$ can be expanded up to a first nonvanishing order such that
$f_A(u)+f_A(-u)\propto \mbox{Tr}A^2 u^2$ and the corresponding
expression coincides with that found by Efetov. As such,
Efetov's formula is indeed  applicable also for  constant
 matrices $A$ of the type described above.

We see that the most striking qualitative features of the Efetov's formula: i)a
nonvanishing density of real eigenvalues and ii) a linear increase with
$|y|$ of the probability density to have a finite imaginary part
-persist for any antisymmetric perturbation $A$.
The strongest quantitative deviation
 from Efetov's result  occurs in
the case of finite-rank perturbations $A$ such that
 $\mu_i=0$ for $i>M$. In particular, one can show that if at least
one of the quantities $g_i=\frac{1}{2\pi\nu}({\mu}_i+{\mu}^{-1}_i)$.
is equal to unity, the mean density decays asymptotically as
$\rho(y\gg 1)\propto y^{-2}$. Such a slow power law decay should be
contrasted with the Gaussian case when  one always has a very
sharp cutoff of the density for large enough $y$. For the general case
of a finite-rank antisymmetric perturbation such that $g_i\ne 1$ the density
is cut exponentially at $y\sim (g_i-1)^{-1}$.

In conclusion, we put forward a conjecture on statistics of
S-matrix poles $Z_i$
for systems with broken time-reversal invariance and
verified that it is perfectly compatible with the existent knowledge on
quantum chaotic scattering. In particular, our conjecture
allowed us to reproduce
 statistical characteristics of Wigner time delays
known from independent calculations.
We analyzed the ensuing two-point statistical measures
as e.g. spectral form factor, number variance and
small distance behavior of the nearest neighbor
distance distribution $p(Z_0,S)$.
In the final part of the paper we calculated
 the density
of complex eigenvalues of an ensemble of real
weakly asymmetric matrices. The expression obtained generalizes the recent
result by Efetov\cite{Efnh}
to the case of an arbitrary antisymmetric perturbation.

 YF is obliged to V.Sokolov and B.Khoruzhenko for comments and
encouragement. The work was supported by SFB 237 "Disorder and Large
Fluctuations", VIII-2 "Russian State Program
for Stat. Physics" and RFBR 96-02-18037a (MT).

\end{document}